\documentclass[prc,showpacs,superscriptaddress,twocolumn]{revtex4}

\usepackage{amsmath,bm}
\usepackage{graphicx}
\usepackage{epstopdf}
\usepackage{subfigure}
\usepackage{epsfig}
\usepackage{amsmath,amssymb,amsfonts}
\usepackage{color}
\usepackage[utf8]{inputenc}
\usepackage{verbatim}
\usepackage{array}
\usepackage{hyperref}

\graphicspath{{figures/}} 

\begin{document}

\title{Production of leptons from decay of heavy-flavor hadrons in high-energy nuclear collisions}

\author{Sa Wang}
\email{wangsa@mails.ccnu.edu.cn}
\affiliation{College of Science, China Three Gorges University, Yichang 443002, China}
\affiliation{Center for Astronomy and Space Sciences and Institute of Modern Physics, China Three Gorges University, Yichang 443002, China}
\affiliation{Key Laboratory of Quark \& Lepton Physics (MOE) and Institute of Particle Physics, Central China Normal University, Wuhan 430079, China}

\author{Yao Li}
\affiliation{Key Laboratory of Quark \& Lepton Physics (MOE) and Institute of Particle Physics, Central China Normal University, Wuhan 430079, China}

\author{Shuwan Shen}
\affiliation{Key Laboratory of Quark \& Lepton Physics (MOE) and Institute of Particle Physics, Central China Normal University, Wuhan 430079, China}

\author{Ben-Wei Zhang}
\email{bwzhang@mail.ccnu.edu.cn}
\affiliation{Key Laboratory of Quark \& Lepton Physics (MOE) and Institute of Particle Physics, Central China Normal University, Wuhan 430079, China}

\author{Enke Wang}
\affiliation{Guangdong-Hong Kong Joint Laboratory of Quantum Matter, Guangdong Provincial Key Laboratory of Nuclear Science, Southern Nuclear Science Computing Center, South China Normal University, Guangzhou 510006, China}

\date{\today}

\begin{abstract}
This paper presents a theoretical study on the production of the heavy-flavour decay lepton (HFL) in high-energy nuclear collisions at the LHC. The pp-baseline is calculated by the FONLL program, which matches the next-to-leading order pQCD calculation with the next-to-leading-log large-$p_T$ resummation. The in-medium propagation of heavy quarks is driven by the modified Langevin equations, which consider both the elastic and inelastic partonic interactions. We propose a method to separate the respective influence of the six factors, such as pp-spectra, the cold nuclear matter (CNM) effects, in-medium energy loss (E-loss), fragmentation functions (FFs), coalescence (Coal), and decay channels, which may contribute to the larger $R_{AA}$ of HFL $\leftarrow b$ compared to that of HFL $\leftarrow c$ in nucleus-nucleus collisions. Based on quantitative analysis, we demonstrate that both coalescence hadronization, decay channels and the mass-dependent E-loss play an essential role at $p_T<5$ GeV, while the latter dominates the higher $p_T$ region. It is also found that the influences of the CNM effects and FFs are insignificant. At the same time, different initial pp-spectra of charm and bottom quarks have a considerable impact at $p_T>5$ GeV. Furthermore, we explore the path-length dependence of jet quenching by comparing the HFL $R_{AA}$ in two different collision systems. Our investigations show smaller HFL $R_{AA}$ in Pb+Pb than in Xe+Xe within the same centrality bin, consistent with the ALICE data. The longer propagation time and more effective energy loss of heavy quarks in Pb+Pb collisions play critical roles in the stronger yield suppression of the HFL compared to that in Xe+Xe. In addition, we observe a scaling behavior of the HFL $R_{AA}$ in Xe+Xe and Pb+Pb collisions.
\end{abstract}


\maketitle

\section{Introduction}
\label{sec:intro}
Over the past few decades, the main goal of the high-energy nuclear collision programs at the Relativistic Heavy Ion Collider (RHIC) and the Large Hadron Collider (LHC) is revealing the mystery of the de-confined nuclear matter, quark-gluon plasma (QGP), at extremely hot and dense condition~\cite{Freedman:1976ub, Shuryak:1977ut}. Investigations on the properties of the QGP are fundamental and essential to test the basic theory of Quantum Chromodynamics (QCD) and understand the phase transition of nuclear matter at high temperature and density~\cite{Connors:2017ptx, Andrews:2018jcm, Cunqueiro:2021wls, Cao:2020wlm}. The strong interactions between the initial-produced high-$p_T$ jet/parton in hard QCD scattering and the thermal particle in the QGP medium, known as the ``jet quenching'' phenomenon, provides a new arena to explore the properties of the QGP ~\cite{Wang:1992qdg, Gyulassy:2003mc, Mehtar-Tani:2013pia, Qin:2015srf, Wang:1998bha, Wang:2001cs, Wang:2002ri, Vitev:2009rd, He:2020iow, Neufeld:2010fj, Dai:2012am, Vitev:2008rz, Zhang:2021sua, Chen:2022kic, Zhang:2021xib}.

Heavy quarks (charm and bottom) have attracted much attention in the community of heavy-ion collision physics owing to their unique advantages \cite{Prino:2016cni, Rapp:2018qla, Dong:2019unq, Zhao:2020jqu, Wang:2023eer, Tang:2020ame}. Firstly, their initial yields in hard QCD processes are perturbatively calculable, resulting from their large mass ($m_Q\gg \Lambda_{QCD}$)~\cite{Andronic:2015wma}, while their thermal production is negligible in the collision energy of current heavy-ion programs~\cite{Zhang:2007yoa}. Secondly, heavy quarks contain their flavour identities as they interact with the thermal quasi-particle in the QGP. In the past decade, plentiful experimental measurements on heavy-flavour hadrons, such as the nuclear modification factor $R_{AA}$~\cite{Adamczyk:2014uip, Adam:2015sza, Sirunyan:2017xss, PHENIX:2011img, STAR:2013eve, ALICE:2014wnc, CMS:2011all, ALICE:2018lyv}, collective flow (direct flow $v_1$~\cite{STAR:2019clv, ALICE:2019sgg}, elliptical flow $v_2$~\cite{Abelev:2014ipa, Adamczyk:2017xur, Acharya:2017qps, Sirunyan:2017plt}) and baryon-to-meson ratio~\cite{STAR:2019ank, Vermunt:2019ecg}, have been extensively made in the high-energy nuclear collisions both at the RHIC and LHC. These efforts greatly facilitate our understanding of the in-medium interaction mechanisms, dynamical correlations, and hadronization patterns of heavy quarks in the strongly-coupled QCD matter \cite{vanHees:2007me, CaronHuot:2008uh, Djordjevic:2015hra, He:2014cla, Chien:2015vja, Kang:2016ofv, Alberico:2013bza, Xu:2015bbz,
 Das:2016cwd, Ke:2018tsh, Xu:2004mz, Altenkort:2020fgs, He:2019vgs, Li:2021nim, Jiang:2022uoe}. Among them, the mass effect of jet quenching is especially a fascinating topic \cite{Dokshitzer:2001zm, Zhang:2003wk, Armesto:2003jh, Djordjevic:2003qk}, which has been deeply explored by comparing the yield suppression of charm- and bottom-hadron in nucleus-nucleus collision relative to p+p in both experiment~\cite{ALICE:2015nvt, CMS:2016mah, CMS:2017uoy, CMS:2017uuv, ATLAS:2018hqe, CMS:2018bwt, ALICE:2022tji} and theory~\cite{Cao:2013ita, Cao:2016gvr, Xing:2019xae, Li:2020umn, Xing:2023ciw}. However, it is still challenging to reconstruct bottom-hadrons in the experiment due to their small cross section and many decay channels \cite{ALICE:2022iba}, which results in significant uncertainties in the current measurements. The lepton from heavy-flavour hadron decays (denoted as HFL, including muon and electron) is viewed as a complementary tool to investigate the mass effect of jet quenching in the QGP. Recent measurements in Au+Au~\cite{PHENIX:2022wim, STAR:2021uzu} and Pb+Pb \cite{ATLAS:2021xtw} collisions show that the yield suppression of the HFL from charm-hadron may be more potent than that from bottom-hadron. Nevertheless, the final-state yield of the HFL in A+A collisions is a complicated interplay of several factors: the initial $p_T$ spectra, initial-state nuclear matter effect, the in-medium energy loss of heavy quarks, fragmentation functions, coalescence probabilities, and decay channels. Investigating how each factor plays a role in the obtained HFL $R_{AA}$ in A+A collisions is crucial, which is essential to capture the mass effect of jet quenching from the current measurements. On the other hand, tremendous efforts have been made to address the system size dependence of jet quenching in different collision systems \cite{Zigic:2018ovr, Shi:2018izg, Huss:2020dwe, Huss:2020whe, Zhang:2022fau}. In this context, the yield suppression of the HFL in Xe+Xe and Pb+Pb collisions will also provide a new opportunity to address the path-length dependence of partonic energy loss. In addition, it is particularly interesting to test the ``scaling behavior'' of the HFL $R_{AA}$ in different collision systems. At the same time, that of heavy-flavour hadrons has been systematically discussed in Ref. \cite{Li:2021xbd}.

This work will investigate the production of the heavy-flavour decay lepton in high-energy nuclear collisions at the LHC. The pp-baseline is provided by the FONLL program, which matches the next-to-leading order pQCD calculation with the next-to-leading-log large-$p_T$ resummation~\cite{Cacciari:1998it, Cacciari:2001td}. The in-medium evolution of heavy quarks is driven by the Langevin transport approach~\cite{Dai:2018mhw, Wang:2019xey, Wang:2020qwe, Wang:2020ukj, Wang:2021jgm, Li:2022tcr} which takes into account the collisional and radiative energy loss \cite{Cao:2013ita, Cao:2014fna, Cao:2015hia}. We will focus on the mass effect and system size dependence of jet quenching by estimating the yield suppression of the HFL in nucleus-nucleus collisions at the LHC. To address the mass effect, we will present a strategy to separate the contributions of several factors that may lead to the different $R_{AA}$ of HFL $\leftarrow c$ and HFL $\leftarrow b$, such as pp-spectra, cold nuclear matter (CNM) effect, in-medium energy loss (E-loss), fragmentation functions (FFs), coalescence and decay channels. To address the system size dependence of jet quenching, we will estimate the averaged energy loss and propagation time of heavy quarks in the two different collision systems, Xe+Xe and Pb+Pb, which may help find the key factor that leads to stronger suppression of the HFL in Pb+Pb compared to that in Xe+Xe. Besides, we will also discuss the scaling behavior of the HFL $R_{AA}$ in the two different collision systems of Xe+Xe and Pb+Pb.

The remainder of this paper is organized as follows. In Sec.~II, the theoretical frameworks used to study the HFL production in nucleus-nucleus collisions will be introduced. In Sec.~III, we will present our main results and conclusions about the mass and path-length dependence of the HFL yield suppression. At last, we will summarize this work in Sec.~IV.

\section{Theoretical framework}
\label{sec:framework}

\begin{figure}[!t]
\begin{center}
\vspace*{0.1in}
\includegraphics[width=3.3in,angle=0]{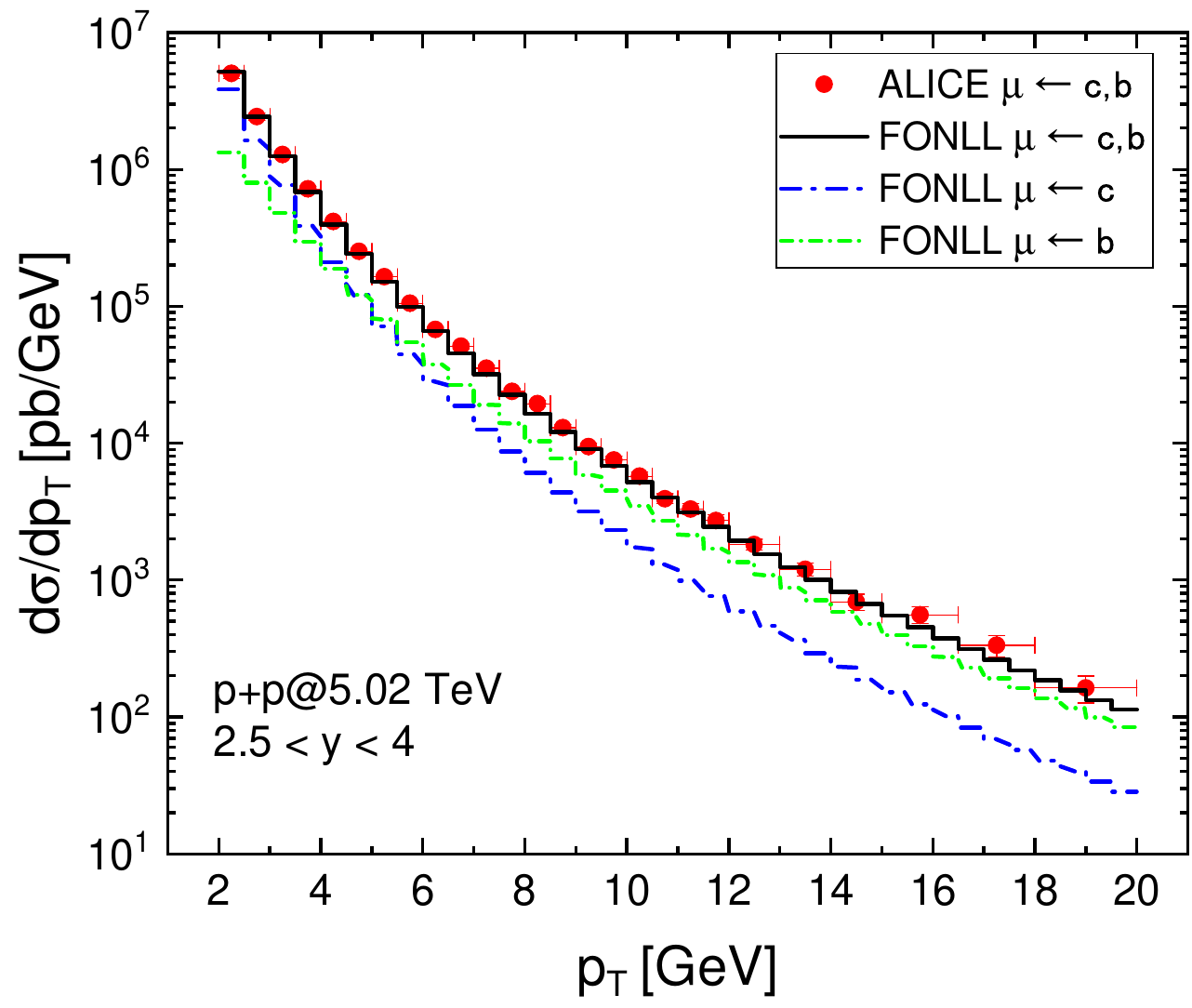}
\vspace*{0.1in}
\caption{Differential cross sections of $\mu\leftarrow c$ (blue dash-dot line), $\mu\leftarrow b$ (green dash-dot line) and $\mu\leftarrow c,b$ (black solid line) versus transverse momentum $p_T$ in p+p collisions at $\sqrt{s}$=5.02 TeV calculated by FONLL, compared with ALICE data (red points)~\cite{ALICE:2019rmo}.}
\label{fig:alipp}
\end{center}
\end{figure}

 The differential cross section ($d\sigma/dp_T$) of heavy quarks in p+p collisions is usually utilized as the baseline to study the yield suppression of the HFL in high-energy nuclear collisions. The higher order corrections and logarithmic-terms resummation play important roles in the heavy quark production in hadron collisions at the LHC \cite{Nason:1987xz, Nason:1989zy, Beenakker:1990maa}. In this work, the initial $p_T$ spectra of charm and bottom quarks are provided by the FONLL scheme~\cite{Cacciari:1998it}, which matches the Fixed Order (FO) next-to-leading order hard QCD processes with the Next-to-Leading-Log (NLL) large-$p_T$ resummation~\cite{Cacciari:2001td}. The FONLL program has been successfully employed to describe the measurement of heavy flavour production at the LHC~\cite{Cacciari:2012ny, Cacciari:2015fta}. As shown in Fig.~\ref{fig:alipp}, the differential cross section of the muon from heavy quarks ( $\mu\leftarrow c,b$ ) calculated by FONLL in p+p collisions at $\sqrt{s}=$ 5.02 TeV is compared to the ALICE data~\cite{ALICE:2019rmo}. It is found that FONLL calculations show a nice agreement with experimental measurements. The contribution from $\mu\leftarrow c$ and $\mu\leftarrow b$ are also estimated in Fig.~\ref{fig:alipp}. We observe that the muon production is dominated by $\mu\leftarrow c$ at $p_T<4$ GeV but by $\mu\leftarrow b$ at high $p_T$. In addition, it should be noted that we take into account the initial cold nuclear matter (CNM) effect in A+A collisions by using the EPPS16 parameterization \cite{Eskola:2016oht} to modify the parton distribution function (PDF) of a free proton, then obtain the $p_T$ spectra of heavy quarks by the FONLL as the input of the calculation of the HFL production in A+A collisions.

In nucleus-nucleus collisions, the evolution of heavy quarks in the strongly coupled nuclear matter can be described by the modified Langevin transport equations~\cite{Cao:2013ita,Cao:2014fna,Cao:2015hia,Dai:2018mhw,Wang:2019xey,Wang:2020qwe,Wang:2020ukj,Wang:2021jgm,Li:2022tcr}.

\begin{eqnarray}
&&\Delta\vec{x}(t)=\frac{\vec{p}(t)}{E}\Delta t\\
&&\Delta\vec{p}(t)=-\eta_D\vec{p}\Delta t+\vec{\xi}(t)\Delta t-\vec{p}_{\rm g}(t)
\label{eq:lang2}
\end{eqnarray}

These two equations correspond to heavy quarks' position and momentum updates during an evolution time $\Delta t$ in the QGP. The three terms on the right-hand side of Eq.~(\ref{eq:lang2}) represent the drag term, the thermal stochastic term, and the momentum recoil term, respectively. Among them, the drag term denotes the energy dissipation as heavy quarks traversing the hot/dense nuclear matter, whose strength is controlled by the drag coefficient $\eta_D$. The second stochastic term denotes the random kicks from quasi-particles in the QGP, and the random force $\vec{\xi}(t)$ is modeled by a Gaussian distribution with mean value zero and variance $\kappa/\Delta t$, where $\kappa$ is the diffusion coefficient in momentum space. These two terms describe the elastic interactions between the heavy quarks and thermal partons in the QGP. Note that $\eta_D$ and $\kappa$ can be correlated by the fluctuation-dissipation theorem $\kappa=2\eta_DET$~\cite{Kubo}. It should also be noted that at very low $p_T$ where the non-perturbative effect is significant, the energy loss of heavy quarks cannot be simply categorized into collisional and radiative contributions. In that background, the Lattice estimation of the heavy quark diffusion coefficient should be helpful for understanding the lower-$p_T$ heavy quark dynamics in heavy-ion collisions, since the Lattice calculations account for all interactions between heavy quarks and the medium even if we cannot list them in perturbative language. Since the medium-induced gluon radiation is significant to the energy loss of heavy quarks at high-$p_T$ region ($p_T^Q>5m_Q$), a correction term $-\vec{p}_{\rm g}$ is effectively considered to describe the momentum decrease caused by such radiation processes \cite{Cao:2013ita}. In our framework, the implementation of in-medium radiation processes of heavy quarks is based on the gluon spectrum calculated with the higher-twist approach~\cite{Guo:2000nz, Zhang:2003yn, Zhang:2003wk, Majumder:2009ge},

\begin{eqnarray}
\frac{dN_g}{ dxdk^{2}_{\perp}dt}=\frac{2\alpha_{s}P(x)\hat{q}}{\pi k^{4}_{\perp}}\sin^2(\frac{t-t_i}{2\tau_f})(\frac{k^2_{\perp}}{k^2_{\perp}+x^2M^2})^4,
\label{eq:dndxk}
\end{eqnarray}
where $x$ and $k_\perp$ are the radiated gluon's energy fraction and transverse momentum. $P(x)$ is the quark splitting function~\cite{Wang:2009qb}, $\tau_f=2Ex(1-x)/(k^2_\perp+x^2M^2)$ the formation time of the daughter gluon. $\hat{q}=q_0(T/T_0)^3p_{\mu}u^{\mu}/E$ denotes the general jet transport parameter in the QGP~\cite{Chen:2010te}, where $T_0$ is the highest temperature in the most central A+A collisions, and $u^{\mu}$ presents flow of the expanding QCD medium. It is reasonable to assume that the gluon radiation is a Poisson process \cite{Cao:2016gvr}, then the probability of radiation during a timestep $\Delta t$ can be calculated as,

\begin{eqnarray}
P(n,t,\Delta t)=\frac{\lambda^n}{n!}e^{-\lambda}
\end{eqnarray}
where n(=0, 1, 2...) denotes the number of radiation at the timestep ($t,t+\Delta t$). $\lambda$ is the average radiation number which can be determined by integrating the spectrum in Eq.~\ref{eq:dndxk}.
\begin{eqnarray}
\lambda=\Delta t\int dx dk^2_{\perp}\frac{dN_g}{dxdk^2_{\perp}dt}
\end{eqnarray}

We have two parameters $\hat{q}$ and $\kappa$ to be determined in the Langevin equations in Eq.~\ref{eq:lang2}. The former has been extracted by fitting the identified hadron production in A+A collisions~\cite{Ma:2018swx}, which gives the best value $q_0=1.2$~GeV$^2$/fm at the LHC. As the $\hat{q}$ is fixed, we obtain $\kappa\sim  \pi T^3$ by a $\chi^2$ fitting to the D meson $R_{AA}$ data~\cite{Sirunyan:2017xss,ALICE:2018lyv}. Note that the $\kappa$ value of our fitting is consistent with the Lattice estimations, including the calculations in a pure gluon plasma \cite{Francis:2015daa} and the quenched results \cite{Brambilla:2020siz, Banerjee:2022gen} ($\kappa=1.8\sim3.4 \, T^3$). The recent reported 2+1 flavour Lattice calculation has shown a smaller value of $\kappa$, ($\kappa(T = 356\, {\rm MeV}) \sim 4.8 \, T^3$), especially at lower temperature ($\kappa(T = 195\, {\rm MeV}) \sim 11 \, T^3$) \cite{Altenkort:2023oms, Altenkort:2023eav}. Since the evolution of heavy quarks near the critical temperature $T_c$ is important for understanding the ``$R_{AA}\otimes v_2$'' puzzle of the heavy flavours measurement \cite{Liao:2008dk, Kopeliovich:2012sc, Noronha-Hostler:2016eow}, it will be of significant interest to systematically explore the differences between phenomenological fitting and the first-principle calculation. Some further and detailed studies have been made to explore the temperature and energy dependence of $\hat{q}$ \cite{JETSCAPE:2021ehl, Kumar:2020wvb, Xie:2022fak} and $\kappa$  \cite{Xu:2017obm, Cao:2018ews, Li:2019lex}. It should be noted that in our current framework, the spatial diffusion coefficient $2\pi TD_s$ is momentum-independent. In future studies, the temperature and momentum dependence of $D_s$ will be taken into account to consider the non-perturbative effects, which play an important role in the interaction between lower $p_T$ heavy quarks and QGP medium. Meanwhile, in order to take into account the detailed balance of gluon absorption and emission \cite{Ke:2018tsh, Xu:2004mz}, a lower cutoff ($E_0=\mu_D=\sqrt{4\pi\alpha_s}T$) has been imposed for the medium-induced gluon radiation \cite{Cao:2013ita, Cao:2014fna, Cao:2015hia}, where $\mu_D$ is the Debye screening mass in the QGP medium. The radiation processes with radiated gluon energy lower than $\mu_D$ are forbidden. Below such a cutoff, the total contributions of the gluon emission and absorption are assumed to be canceled. As a phenomenological treatment of the detailed balance of the gluon emission and absorption of heavy quarks in the QGP, it effectively guarantees that the heavy quark can reach the thermal equilibrium for enough propagation time.

In this study, the time-space evolution of the hot QCD medium is described by the 2+1D CLVisc hydrodynamic model~\cite{Pang:2016igs, Pang:2018zzo, Wu:2021fjf}, which provides information on the temperature and velocity of the medium cells. The starting time $\tau_0=$ 0.6 fm, the shear viscosity to entropy ratio $\eta_{v}/s=$ 0.08, and the freeze-out temperature $T_{f}=$ 137 Mev are used in the calculations of the hydro profile, which are consistent with the setup in Ref. \cite{Pang:2018zzo}. The initial entropy density distribution $s(\tau_0, x, y)$ is provided by the Trento program \cite{Moreland:2014oya} based on the Glauber model~\cite{Miller:2007ri}, in which the nuclear densities of Pb and Xe obey the Woods-Saxon distribution~\cite{ALICE:2018cpu, Sievert:2019zjr},

\begin{eqnarray}
\rho(r,\theta)=\frac{\rho_0}{1+{\rm exp}[\frac{r-R(\theta)}{a}]}
\label{eq:Woods-Saxon}
\end{eqnarray}
where $\rho_0$ is the nuclear saturation density, $a$ the nuclear skin thickness. $R(\theta)=R_0(1+\beta_2 Y_{20}(\theta))$ is the nuclear radius, where $R_0\sim A^{1/3}$ is the average radius, and the spherical harmonic function $Y_{20}(\theta)$ describes the deformation of an axially symmetric nucleus. The parameter setups of Xe and Pb are listed in Tab.~\ref{tab:WS}.

\begin{table}[!t]
\begin{center}
\vspace*{0.2in}
\hspace*{0in}
\begin{tabular}{|p{2.0cm}<{\centering}|p{2.0cm}<{\centering}|p{2.0cm}<{\centering}|p{2.0cm}<{\centering}|}
  \hline
  Nucleus & a [fm] & $R_0$ [fm] & $\beta_2$ \\
  \hline
  $^{129}$Xe & 0.590 & 5.40 & 0.18 \\
  \hline
  $^{208}$Pb & 0.546 & 6.62 & 0 \\
  \hline
\end{tabular}
\caption{Parameter setup of Woods-Saxon distributions of Xe and Pb in the Glauber calculations \cite{ALICE:2018cpu,Sievert:2019zjr}.}
\end{center}
\label{tab:WS}
\end{table}

Heavy quarks propagate and lose energy in the QGP with Eq.~\ref{eq:lang2} until the local medium reaches a critical temperature. After the in-medium evolution, the heavy quarks will hadronize into heavy-flavour meson or baryon by fragmentation and coalescence mechanisms \cite{Altmann:2024kwx, He:2019vgs, Song:2015sfa, Plumari:2017ntm, Beraudo:2017gxw}. It should be noted that the coalescence of heavy and light quarks is the dominant mechanism of hadronization at low $p_T$. We utilize the ``sudden recombination'' approach to consider the production of heavy-flavour hadron by coalescence \cite{Cao:2013ita, Oh:2009zj}, in which the momentum distribution of produced heavy-flavour meson and baryon can be expressed as:

\begin{eqnarray}
\frac{dN_{M}}{d^3p_{M}}=& &\int d^3p_1d^3p_2 \frac{dN_1}{d^3p_2}\frac{dN_2}{d^3p_2}f^W_M(\vec{p}_1,\vec{p}_2)\delta(\vec{p}_M-\vec{p}_1-\vec{p}_2) \nonumber \\
\\
\frac{dN_{B}}{d^3p_{B}}=& &\int d^3p_1d^3p_2d^3p_3 \frac{dN_1}{d^3p_1}\frac{dN_2}{d^3p_2}\frac{dN_3}{d^3p_3}f^W_B(\vec{p}_1,\vec{p}_2,\vec{p}_3) \nonumber \\
& &\times\delta(\vec{p_B}-\vec{p}_1-\vec{p}_2-\vec{p}_3)
\label{eq:dnq}
\end{eqnarray}
where $dN_i/d^3p_i$ denotes the momentum distribution of the $i$-th valence quark in the heavy-flavour meson or baryon. For light quarks with spin-color degeneracy $g_q$, it is convenient to take the Fermi-Dirac distribution $g_qV/(e^{E/T_c}+1)$ in the local medium cell with volume $V$. As for heavy quark, $dN_i/d^3p_i$ represents the momentum distribution after the in-medium propagation. $f^W_M$ and $f^W_B$ are the Wigner functions describing the probabilities of light and heavy quarks to combine into heavy-flavour meson and baryon, respectively. By assuming that the hadron has a simple harmonic oscillator wave function and averages over the position space, the Wigner functions of heavy-flavour meson may have the following form,

\begin{eqnarray}
f_M^W(q^2)&=&Ng_M\frac{(2\sqrt{\pi}\sigma)^3}{V}e^{-q^2\sigma^2}
\label{eq:wigner1}
\end{eqnarray}
where $q$ represents the relative momenta of quarks in the center of mass frame of the formed meson. The width parameter $\sigma$ is related to the oscillator frequency $\omega$ and the reduced mass $\mu$. The values of the parameters we employed are consistent with that in Ref.~\cite{Cao:2013ita}. Note that Eq.~\ref{eq:wigner1} can be generalized to the case of heavy-flavour baryon by combining with another valence light quark.

\begin{eqnarray}
f_B^W(q_1^2,q_2^2)&=&Ng_B\frac{(4\pi\sigma_1\sigma_2)^3}{V^2}e^{-q_1^2\sigma_1^2-q_2^2\sigma_2^2}
\label{eq:wigner2}
\end{eqnarray}
where $q_1$ and $\sigma_1$ are the relative momentum and width parameters of the first two quarks. $q_2$ and $\sigma_2$ can be viewed as the momentum and width parameters of the third quark in the center-of-mass frame of the first two-particle system. $g_M$ and $g_B$ are the spin-color degeneracy of heavy-flavour meson and baryon. In this work we have taken into account both the ground states of heavy flavor hadron ($D^0, D^+, D^+_s,$ $\Lambda_c^+, B^0, B^+,$ $B^0_s, \Lambda_b^0, \Sigma_Q, \Xi_Q, \Omega_Q$) and their first excited states ($D^{*0}, D^{*+}, D^{*+}_s, B^{*0}, B^{*+}, B^{*0}_s, \Sigma^*_Q, \Xi^*_Q, \Omega^*_Q$). For a heavy quark with fixed momentum $p_Q$, one can integrate Eq.~\ref{eq:wigner1} and Eq.~\ref{eq:wigner2} then obtain the total coalescence probability $P_{\rm coal}(p_Q)$. Note that the normalization factor $N$ can be determined by requiring $P_{\rm coal}(p_Q=0)=1$.

In the realistic simulation, firstly, one can judge whether the coalescence (meson or baryon) occurs by comparing the coalescence probability $P_{coal}$ with a random number between 0 and 1, $R(0,1)$. Secondly, if the coalescence occurs, the prior four-momentum of valence light quarks are sampled by the Fermi-Dirac distribution. Thirdly, one can further judge whether such prior four-momentum of valence light quarks can be accepted using the Rejection Sampling method based on the Eq.~\ref{eq:wigner1} and Eq.~\ref{eq:wigner2}. If it is rejected, one can repeat the sampling of four-momentum of valence light quarks until accepted. Note that for the coalescence of heavy-flavour baryon, the four-momentum of the two thermal valence light quarks can be obtained by the above procedures one by one. If the coalescence does not occur, the heavy quark will fragment into heavy-flavour hadrons. Finally, the semileptonic decay of all the heavy-flavour hadron produced into a lepton is implemented by the ParticleDecays module of PYTHIA8~\cite{Sjostrand:2014zea}.

\section{Results and discussions}
\label{sec:results}

\begin{figure}[!t]
\begin{center}
\vspace*{0.2in}
\includegraphics[width=3.3in,angle=0]{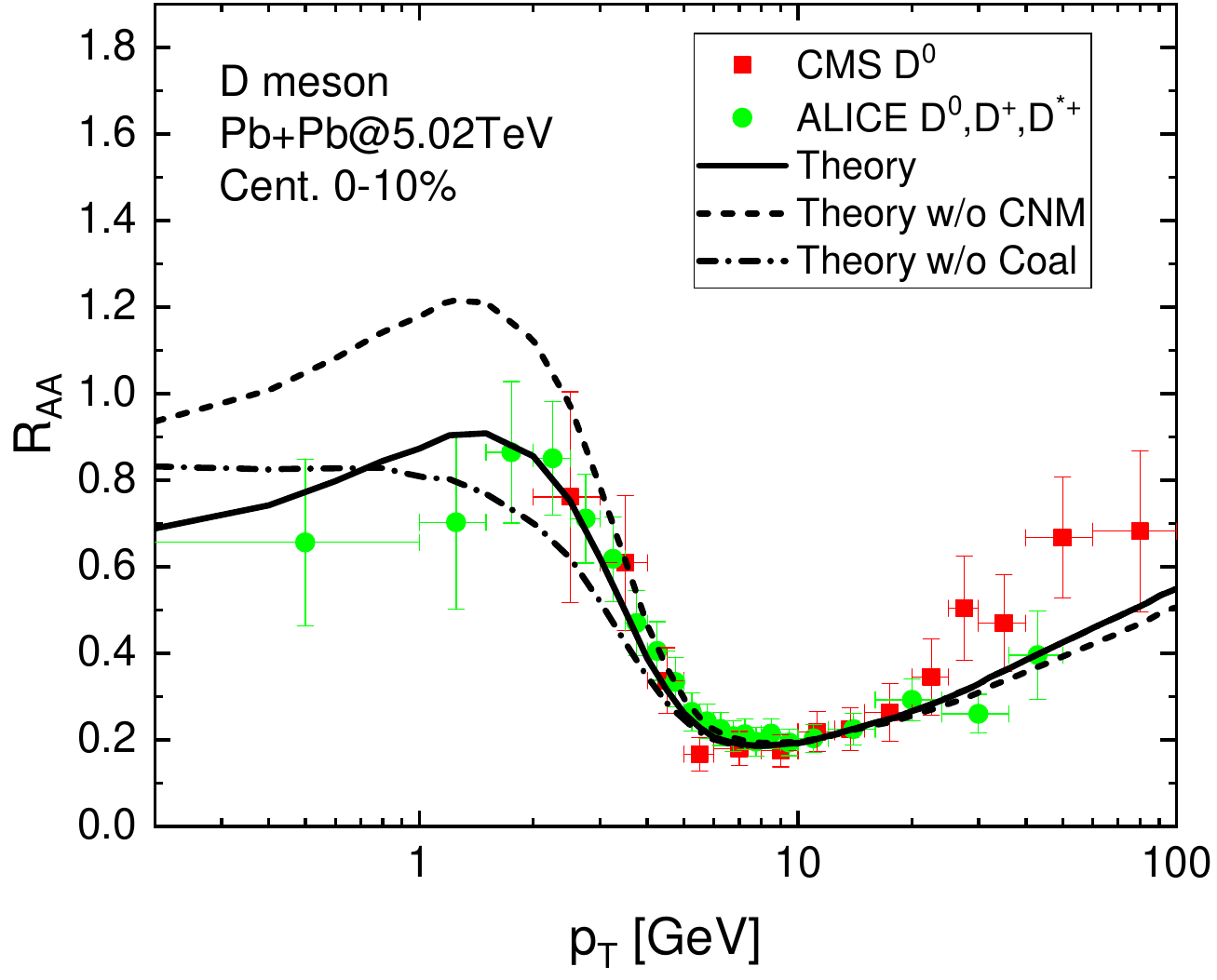}
\includegraphics[width=3.3in,angle=0]{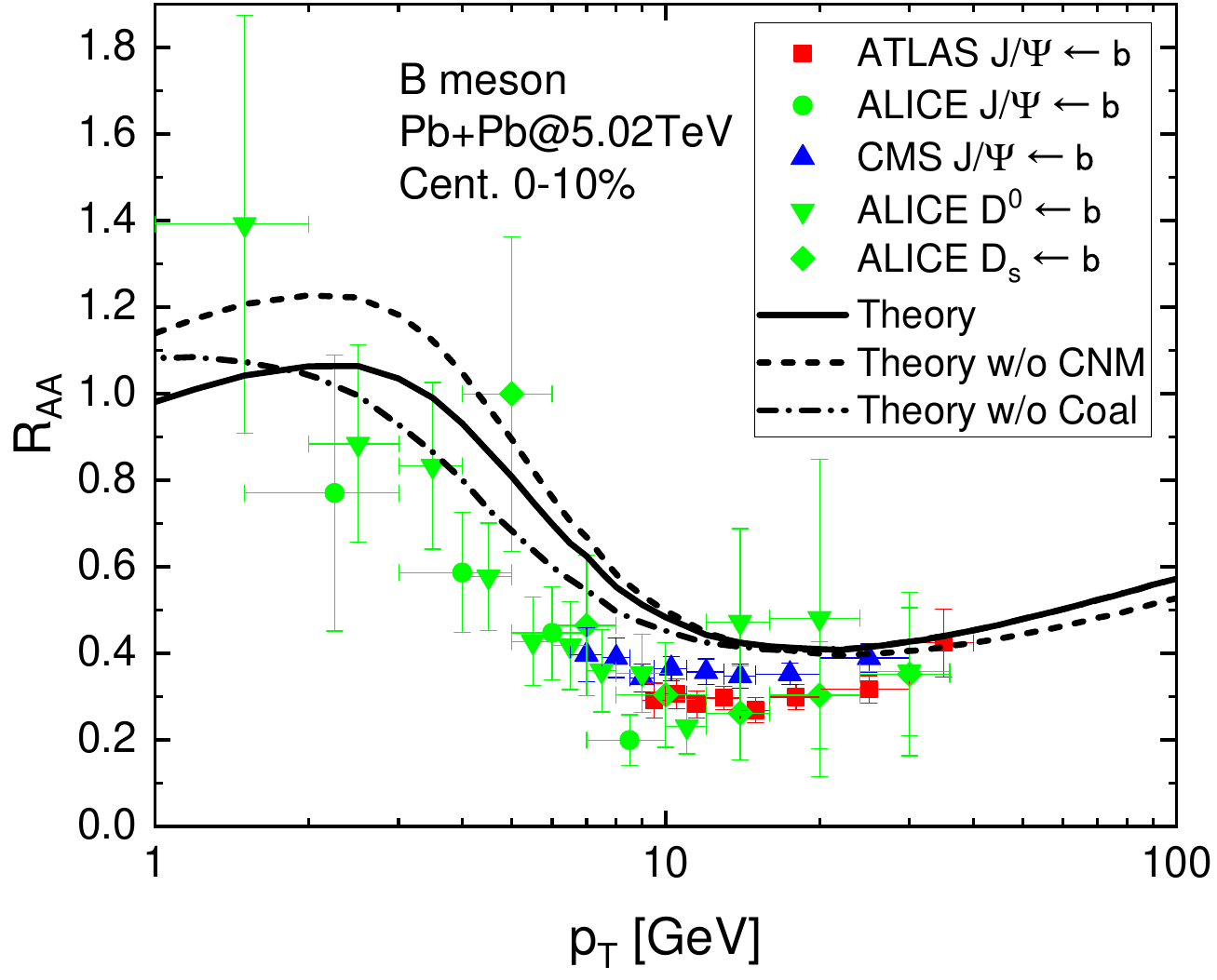}
\vspace*{0.1in}
\caption{The model calculated $R_{AA}$ of D meson (upper panel) and B meson (lower panel) versus $p_T$ in $0-10\%$ Pb+Pb collision at $\sqrt{s_{NN}}=$ 5.02 TeV compared to the ALICE \cite{ALICE:2018lyv, ALICE:2022xrg, ALICE:2023hou, ALICE:2022tji}, ATLAS \cite{ATLAS:2018hqe} and CMS~\cite{Sirunyan:2017xss, CMS:2017uuv} data.}
\label{fig:RAA-DB}
\end{center}
\end{figure}

\begin{figure}[!t]
\begin{center}
\vspace*{0.1in}
\includegraphics[width=3.3in,angle=0]{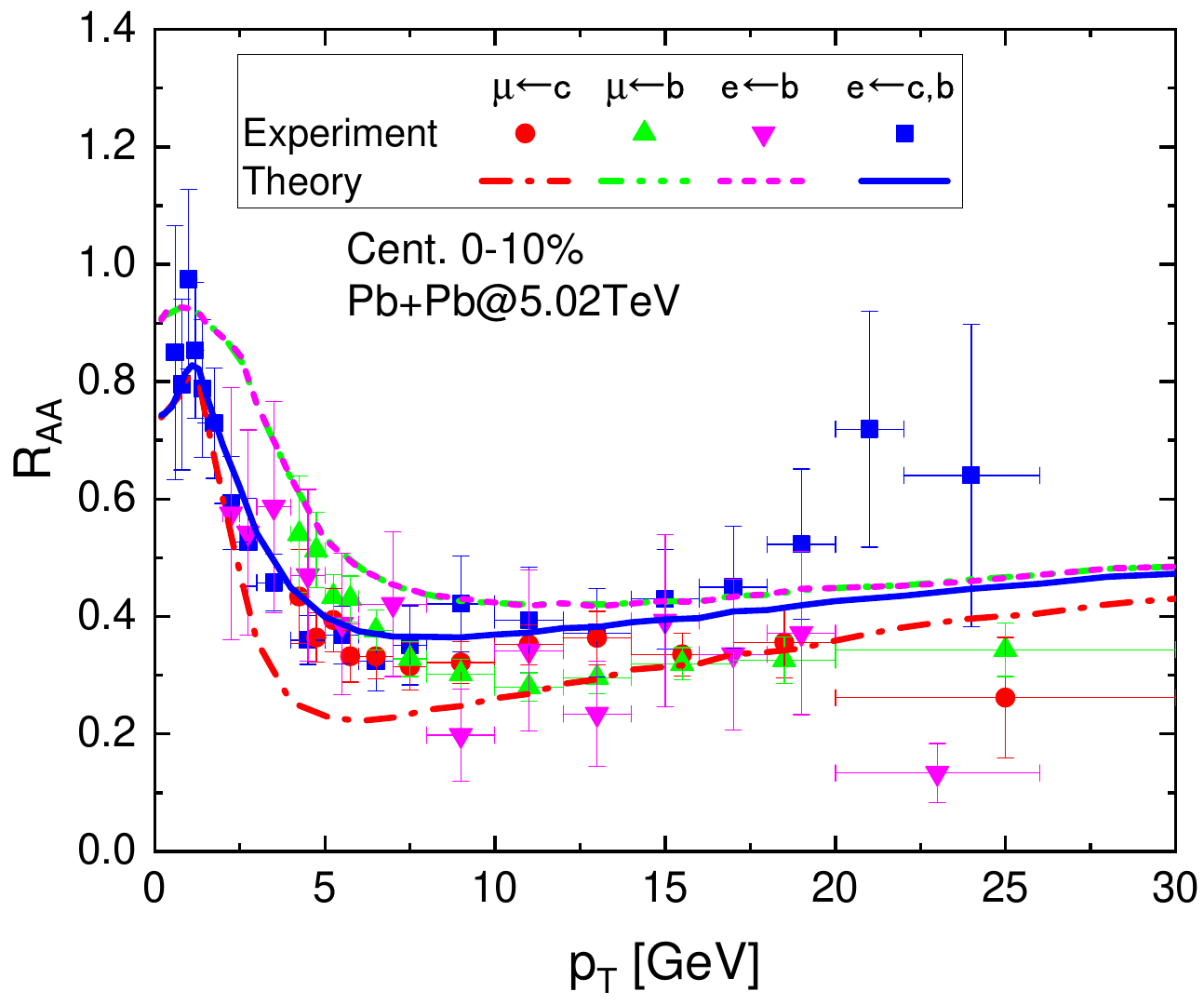}
\includegraphics[width=3.3in,angle=0]{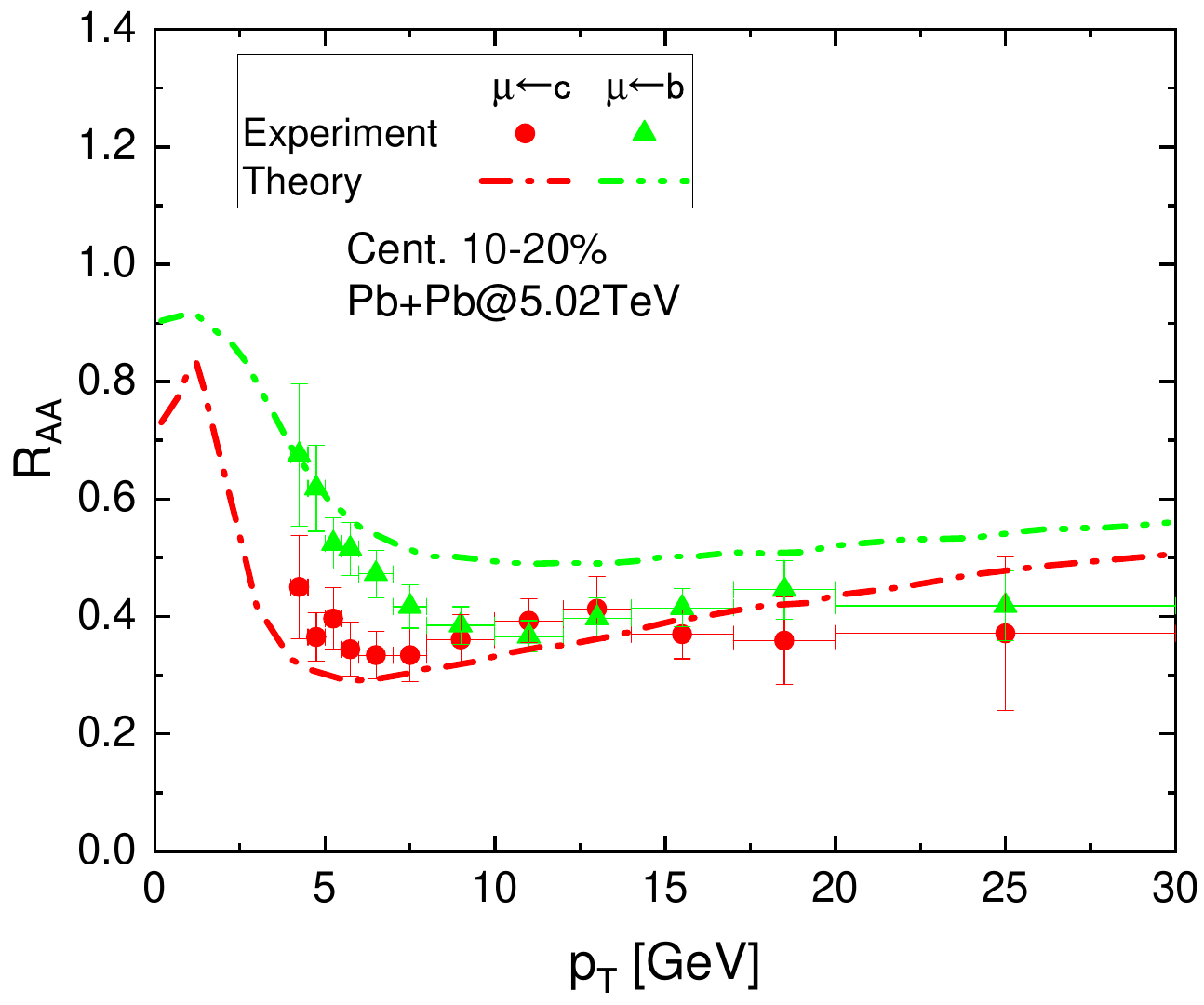}
\vspace*{-0.03in}
\caption{The model calculated HFL $R_{AA}$ as a function of $p_T$ in $0-10\%$ (upper panel) and $10-20\%$ (lower panel) Pb+Pb collision at $\sqrt{s_{NN}}=$ 5.02 TeV compared to the ALICE \cite{ALICE:2019nuy,ALICE:2022iba} and ATLAS \cite{ATLAS:2021xtw} data.}
\label{fig:hfl-pb}
\end{center}
\end{figure}

\subsection{Mass dependence of the HFL yield suppression in Pb+Pb collisions}
In this section, we present the calculations and discussions on the HFL production in high-energy nuclear collisions at the LHC to study the mass dependence of jet quenching. Firstly, to calibrate our model calculation, we show the $R_{AA}$ of charm and bottom hadrons and their lepton in nucleus-nucleus collisions compared with the available experimental data at the LHC. Furthermore, since the final obtained HFL $R_{AA}$ depends not only on the in-medium energy loss but also on other factors, such as the pp-spectra, the CNM effect, fragmentation functions, coalescence hadronization, and decay channel, which may also result in the difference of $R_{AA}$ between HFL $\leftarrow c$ and HFL $\leftarrow b$. To extract the net mass effect of E-loss reflected in the HFL $R_{AA}$, we will propose a strategy of the test particle to estimate the sensitivity of the HFL $R_{AA}$ to those six factors.

As shown in Fig.~\ref{fig:RAA-DB}, we present the calculated $R_{AA}$ of D meson (upper panel) and B meson (lower panel) as compared to the available experimental data measured by the ALICE \cite{ALICE:2018lyv, ALICE:2022xrg, ALICE:2023hou, ALICE:2022tji}, ATLAS \cite{ATLAS:2018hqe} and CMS~\cite{Sirunyan:2017xss, CMS:2017uuv} collaborations, where the nuclear modification factor $R_{AA}$ quantifies the yield suppression of the particle in A+A relative to p+p, and is usually defined as follows,

\begin{eqnarray}
R_{AA}(p_T)=\frac{1}{\left\langle N_{\rm coll}\right\rangle}\frac{dN^{\rm AA}/dp_{\rm T}}{dN^{\rm pp}/dp_{\rm T}}
\label{eq:raa}
\end{eqnarray}
where $\left\langle N_{\rm coll}\right\rangle$ represents the average number of binary nucleon-nucleon collisions per A+A event. $dN^{\rm AA}/dp_{\rm T}$ and $dN^{\rm pp}/dp_{\rm T}$ are the $p_T$ spectra of the particles in A+A and p+p collisions respectively. The theoretical calculations of $D$ meson $R_{AA}$ are compared with the measurement of prompt charm hadron, including $D^0$, $D^+$ and $D^{*+}$. The theoretical calculations of $B$ meson $R_{AA}$ are also compared with the measurement of non-prompt charm hadron, including $D^0$, $J/\Psi$ and $D_s$. As one can see, the theoretical calculation can well describe the D meson $R_{AA}$ while slightly overestimating the yield suppression of bottom quarks at the intermediate $p_T$. In these calculations, we have also tested the influences of the CNM effect and the coalescence hadronization. Moreover, we find that coalescence hadronization can lead to a ``bump'' structure of D meson $R_{AA}$ near $1\sim3$ GeV, which is critical for the dynamics of low-$p_T$ heavy quarks in the QGP medium. One can also observe that the influence of coalescence for the B meson can expand to a broader $p_T$ region compared to the D meson because the coalescence probability of the bottom quark decreases more slowly with momentum compared to charm. In addition, taking into account the CNM effect is also important for the yield suppression of D meson at $p_T<10$ GeV. The influences of these two factors gradually diminishes at higher $p_T$.

In the upper panel of Fig.~\ref{fig:hfl-pb}, we also present the calculated HFL $R_{AA}$ in central $0-10\%$ Pb+Pb collisions at $\sqrt{s_{NN}}=$ 5.02 TeV compared to the available experimental data \cite{ALICE:2019nuy,ATLAS:2021xtw,ALICE:2022iba}. Firstly, we show the theoretical results of $e\leftarrow c,b$ $R_{AA}$ (solid blue line) to the ALICE data \cite{ALICE:2019nuy} (blue square points) which show a good agreement. In addition, to address the mass dependence of the HFL yield suppression, we also estimate the $R_{AA}$ of $e\leftarrow c$, $e\leftarrow b$ and $\mu\leftarrow b$, as well as the comparison to the ATLAS \cite{ATLAS:2021xtw} and ALICE data \cite{ALICE:2022iba} respectively. For HFL $\leftarrow b$, our $R_{AA}$ calculations generally agree with the experiment data. We observe no difference in the yield suppression between the electron and muon, which is consistent with the ATLAS measurements. For $e\leftarrow c$, our theoretical results appear to underestimate the $R_{AA}$ at $p_T<$ 10 GeV. We find that the $R_{AA}$ of HFL $\leftarrow c,b$ is close to that of HFL $\leftarrow c$ at low $p_T$ but similar to that of HFL $\leftarrow b$ at high $p_T$. It is because the HFL production is dominated by HFL $\leftarrow c$ at low $p_T$ while by HFL $\leftarrow b$ at high $p_T$ as shown in Fig.~\ref{fig:alipp}. In the most central ($0-10\%$) collisions, the experimental measurements of $R_{AA}$ do not show much difference between $\mu\leftarrow c$ and $\mu\leftarrow b$ as our model predicted within uncertainties. However, as shown in the lower plot of Fig.~\ref{fig:hfl-pb}, we also notice that in $10-20\%$ centrality bin, the ATLAS data exhibits a visible distinction between $\mu\leftarrow c$ and $\mu\leftarrow b$ at $p_T<$ 10 GeV, which is well captured by our calculations. Therefore, further experimental efforts should clarify the potential differences of the HFL $R_{AA}$ from the charm- and bottom-hadron decay at the LHC. In addition, we have also checked that the influence of coalescence mechanism on heavy-flavour decay lepton is moderate compared to that manifested in heavy-flavour meson (shown in Fig.~\ref{fig:RAA-DB}) and exits at lower $p_T$ region. It is because the lower-$p_T$ lepton is decayed from the higher-$p_T$ meson. The region where the coalescence mechanism has important influences on the meson $R_{AA}$ will be naturally scaled to lower $p_T$ for lepton.

An important topic in heavy-ion collisions is the ``dead-cone'' effect~\cite{Dokshitzer:2001zm, ALICE:2021aqk}, which suppresses the probability of the medium-induced gluon radiation of a massive quark within a small cone ($\theta<\frac{m}{E}$). Since the ``dead-cone'' effect is closely related to the quark mass, a heavier quark is expected to lose less energy than a relatively lighter one in the QGP, which leads to the mass hierarchy of partonic energy loss $\Delta E_q>\Delta E_c>\Delta E_b$. The higher $R_{AA}$ of $e\leftarrow b$ relative to $e\leftarrow c$ measured by the ATLAS Collaboration \cite{ATLAS:2021xtw} as shown in the lower panel of Fig.~\ref{fig:hfl-pb} is consistent with our expectation, which may hint $\Delta E_c>\Delta E_b$. However, besides the different mass effect of in-medium E-loss between charm and bottom quarks, one should keep in mind that the final obtained lepton $R_{AA}$ of $e\leftarrow b$ and $e\leftarrow c$ also depend on other factors, e.g., pp-spectra, the cold nuclear matter effect, fragmentation functions, coalescence hadronization, and semileptonic decay channels. In other words, even without the mass dependence of in-medium E-loss, the $R_{AA}$ of $e\leftarrow b$ and $e\leftarrow c$ should not be the same. Therefore, it is crucial to understand how each of these factors influences the HFL $R_{AA}$ in nucleus-nucleus collisions.

In this work, we present a strategy to separate the respective contributions of these six factors (denoted as pp-spectra, CNM, E-loss, FFs, Coal, and Decay). To quantify their respective contribution, we will use the test particle (TP) method to estimate the influence of each factor by comparing it with the calculation of realistic charm quarks. Specifically, different from the $R_{AA}$ calculations of HFL $\leftarrow c$, we implement the calculations of the test particles for six cases, in which the treatment of one of the six factors is replaced by that of bottom quarks and the rest five are kept the same as that of charm. The details of this strategy are introduced as follows.

\begin{itemize}
\item {\bf Case-1}: To estimate the influence of the {\bf pp-spectra}, the initial transverse momentum of the test particles is sampled with the pp-spectra of bottom quarks calculated by FONLL. The rest of the treatments are kept the same as that of HFL $\leftarrow c$.
\item {\bf Case-2}: Since the CNM effects in A+A collisions are different for the charm and bottom quarks, to estimate the difference caused by the initial {\bf CNM} effect, the shadowing effects of bottom quarks are considered for the test particles. The rest of the treatments are kept the same as that of HFL $\leftarrow c$.
\item {\bf Case-3}: To estimate the influence of the mass-dependent {\bf E-loss} of heavy quarks, the test particles are assigned with the bottom quark mass provisionally when they propagate in the QGP medium. The rest of the treatments are kept the same as that of HFL $\leftarrow c$.
\item {\bf Case-4}: Since the hadronization of bottom quarks is usually with harder {\bf FFs} with respect to that of charm, to estimate the difference caused by the different FFs, the test particles will fragment into hadron by using parameter setup of the bottom quarks ($\epsilon=0.005$) in Peterson FFs. The rest of the treatments are kept the same as that of HFL $\leftarrow c$. In this case, we switch off the coalescence hadronization.
\item {\bf Case-5}: To estimate the influence of the different coalescence hadronization of charm and bottom quarks to HFL $R_{AA}$, the test particles will hadronize by fragmentation or coalescence mechanism by the setup of bottom quarks, denoted as {\bf FFs+Coal}. The rest of the treatments are kept the same as that of HFL $\leftarrow c$. In this case, we switch on the coalescence hadronization.
\item {\bf Case-6}: The different {\bf Decay} channels of $D\rightarrow $HFL and $B\rightarrow$HFL may also lead to the difference in the HFL $R_{AA}$, the D mesons fragmented from the test particles will mimic as B meson (by changing their particle ID provisionally) to decay into the HFL within PYTHIA8. The rest of the treatments are kept the same as that of HFL $\leftarrow c$.
\end{itemize}

\begin{figure}[!t]
\begin{center}
\vspace*{0.0in}
\includegraphics[width=3.3in,angle=0]{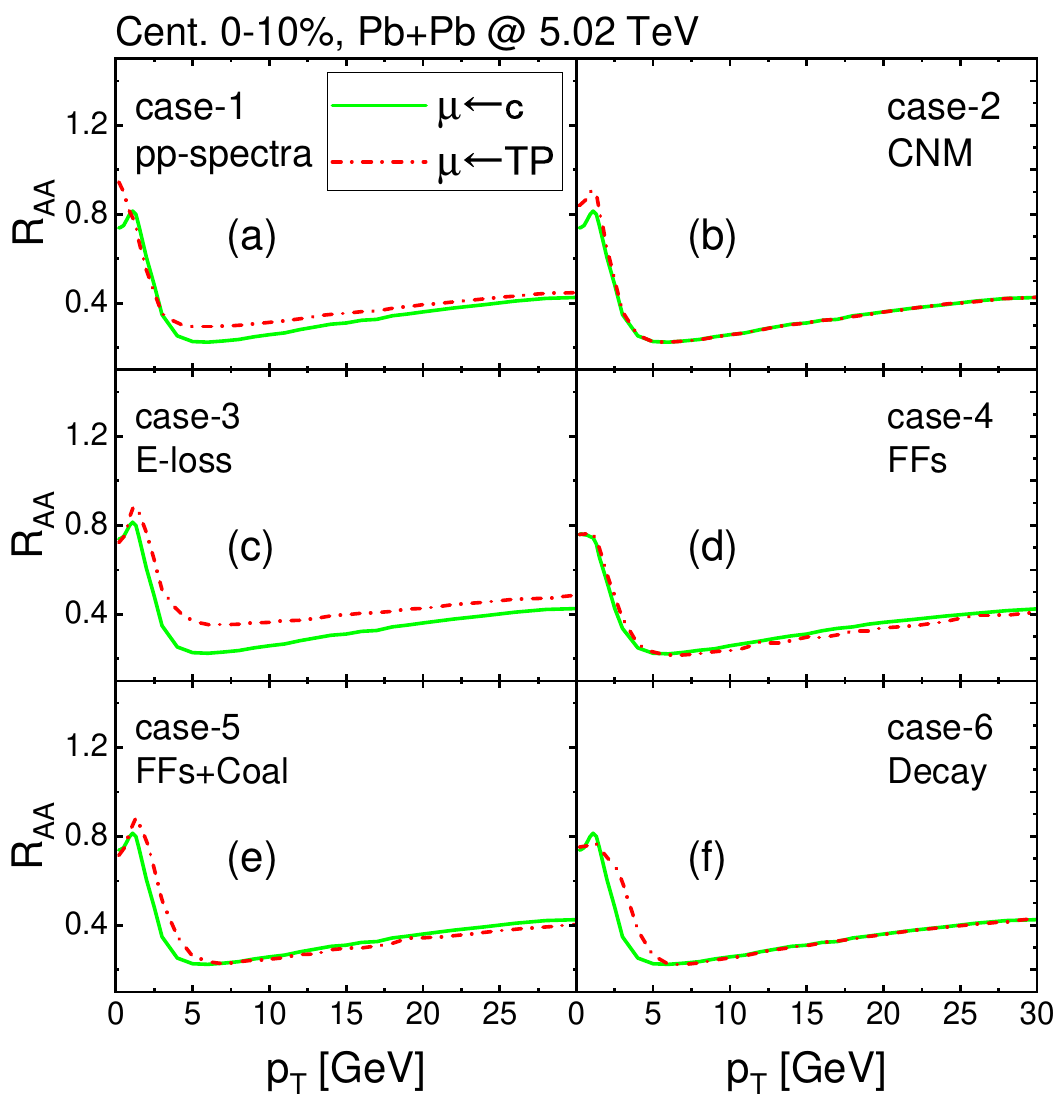}
\vspace*{0.1in}
\caption{The comparisons of the calculated $R_{AA}$ between $\mu\leftarrow$ TP and $\mu\leftarrow c$ for the six cases (a-f).}
\label{fig:5factor}
\end{center}
\end{figure}

By comparing the $R_{AA}$ of the HFL decay from the test particles (HFL $\leftarrow$ TP) for these six cases with that of HFL $\leftarrow c$, one can obtain the difference caused by the variation of each factor. In Fig.~\ref{fig:5factor}, we show the calculated muon $R_{AA}$ decay from the test particles (TP) compared to the realistic case of charm quarks in central $0-10\%$ Pb+Pb collisions at $\sqrt{s_{NN}}=5.02$ TeV. Note that the differences between the two $R_{AA}$ curves of HFL $\leftarrow$ TP and HFL $\leftarrow c$ quantify the influence of the respective factor. The diagram {\bf(a)} represents the influence from the choice of initial pp-spectra, in which we can observe visible larger $R_{AA}$ of $\mu\leftarrow $ TP at $3<p_T<10$ GeV compared to that of $\mu\leftarrow c$. It indicates that the different initial spectra of bottom and charm quarks can result in larger $R_{AA}$ of $\mu\leftarrow b$ relative to $\mu\leftarrow c$. The diagram {\bf(b)} represents the difference caused by the CNM effect suffered in the initial production of charm and bottom quarks in nucleus-nucleus collisions. The muon $R_{AA}$ has almost no difference, only slightly larger values of $\mu\leftarrow $ TP compared to $\mu\leftarrow c$ at $p_T<$ 2 GeV. The diagram {\bf(c)} shows the $R_{AA}$ difference only caused by the mass-dependent energy loss of heavy quarks. It is found that the mass effect of in-medium energy loss leads to larger $R_{AA}$ of $\mu\leftarrow $ TP compared to $\mu\leftarrow c$ at a vast $p_T$ region of 2 to 30 GeV. The diagram {\bf(d)} represents the $R_{AA}$ difference caused by the FFs choices. Besides the moderately larger $R_{AA}$ of $\mu\leftarrow $ TP compared to $\mu\leftarrow c$ at $p_T<5$ GeV, we observe that using FFs of bottom quarks can lead to slightly smaller $R_{AA}$ at higher $p_T$. The diagram {\bf(e)} shows the $R_{AA}$ difference caused by the combined effect of fragmentation and coalescence during hadronization in $\mu\leftarrow c$ and $\mu\leftarrow b$. Compared to Case-4, considering the coalescence mechanism leads to evident larger $R_{AA}$ of test particle than charm quarks at $p_T<5$ GeV. The larger coalescence probability of bottom quarks will contribute to the larger $R_{AA}$ of $\mu\leftarrow b$ compared to $\mu\leftarrow c$ at lower $p_T$. The diagram {\bf(f)} shows the $R_{AA}$ difference caused by different decay channels of $\mu\leftarrow c$ and $\mu\leftarrow b$. We observe that using the bottom-hadron's decay channel, the HFL from test particles has a considerably larger $R_{AA}$ compared to that from charm quarks. It suggests that the different decay channels of the bottom- and charm-hadron may be critical to the larger $R_{AA}$ of $\mu\leftarrow b$ than $\mu\leftarrow c$ at lower $p_T$.

\begin{figure}[!t]
\begin{center}
\vspace*{0.2in}
\includegraphics[width=3.3in,angle=0]{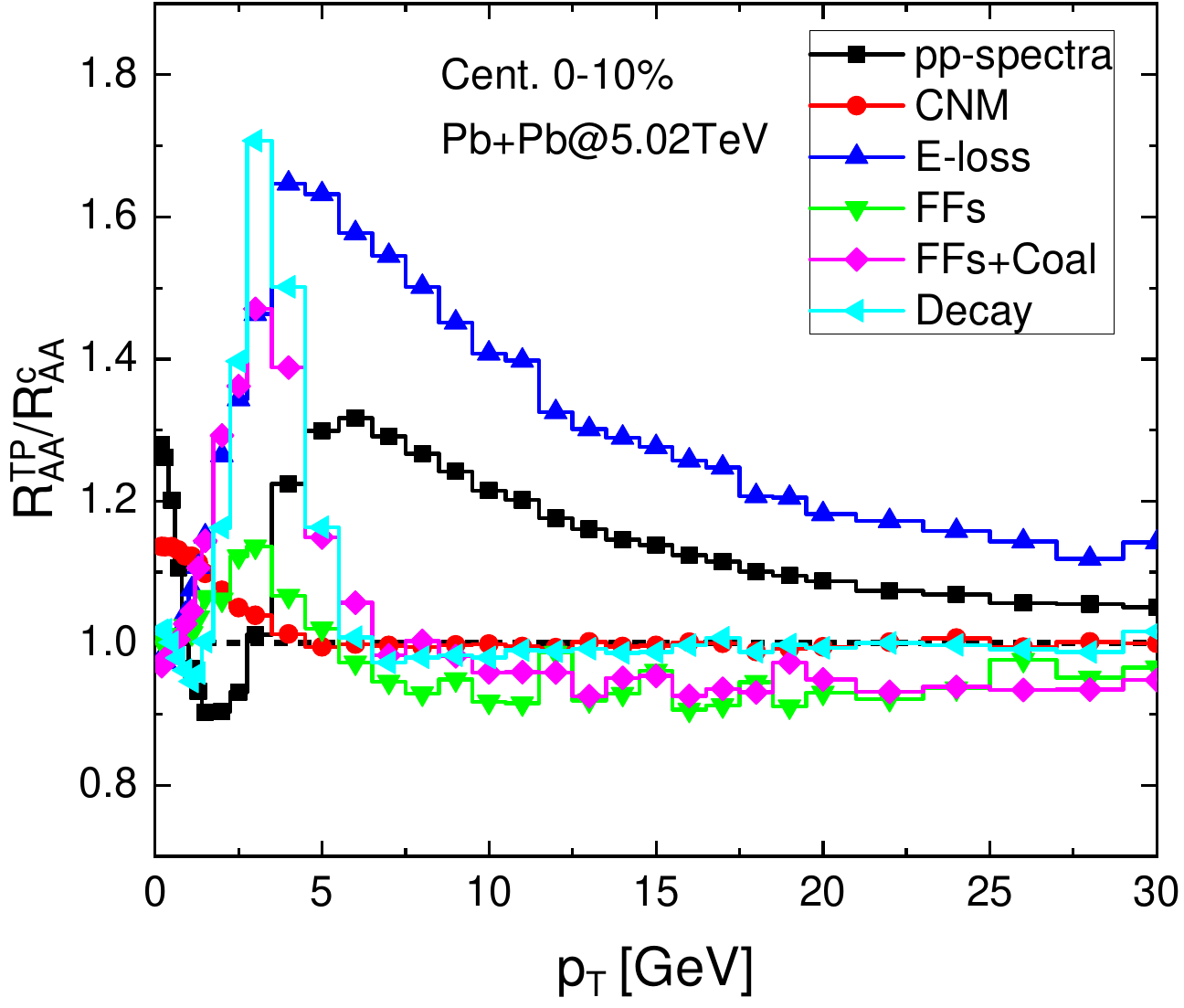}
\vspace*{0.1in}
\caption{Ratios of $R_{AA}$ of $\mu\leftarrow $ TP to $\mu\leftarrow c$ versus $p_T$ in $0-10\%$ Pb+Pb collision at $\sqrt{s_{NN}}=$ 5.02 TeV for the six cases, case-1 (circle line), case-2 (up-triangle line), case-3 (down-triangle line), case-4 (rhombus line), case-5 (left-triangle line), case-6 (square line).}
\label{fig:5factor-r}
\end{center}
\end{figure}

As shown in Fig.~\ref{fig:5factor-r}, we plot the ratios of $R_{AA}$ of $\mu\leftarrow $ TP to that of $\mu\leftarrow c$ for the six cases as a function of $p_T$ in central $0-10\%$ Pb+Pb collisions at $\sqrt{s_{NN}}=$ 5.02 TeV. As we can see, at $p_T<5$ GeV, the mass effect, different coalescence hadronization and decay channel are the important factors that leads to the larger $R_{AA}$ of $\mu\leftarrow $ TP compared to $\mu\leftarrow c$. We find that the mass effect of E-loss of heavy quarks dominates the higher $p_T$ region. It is noted that the pp-spectra also makes a considerable contribution at $p_T>5$ GeV, which remains nonzero even at high $p_T$. In addition, the FFs have considerable influence only at $p_T<5$ GeV, while the influence of CNM effects is minimal. Based on these investigations, we can now conclude that at lower $p_T$ both the mass effect, coalescence hadronization and decay channel of bottom quarks play important roles in the larger $R_{AA}$ of $\mu\leftarrow b$ compared to that of $\mu\leftarrow c$ while the mass-dependent E-loss becomes the key factor at higher $p_T$.

\begin{figure}[!t]
\begin{center}
\includegraphics[width=3.3in,angle=0]{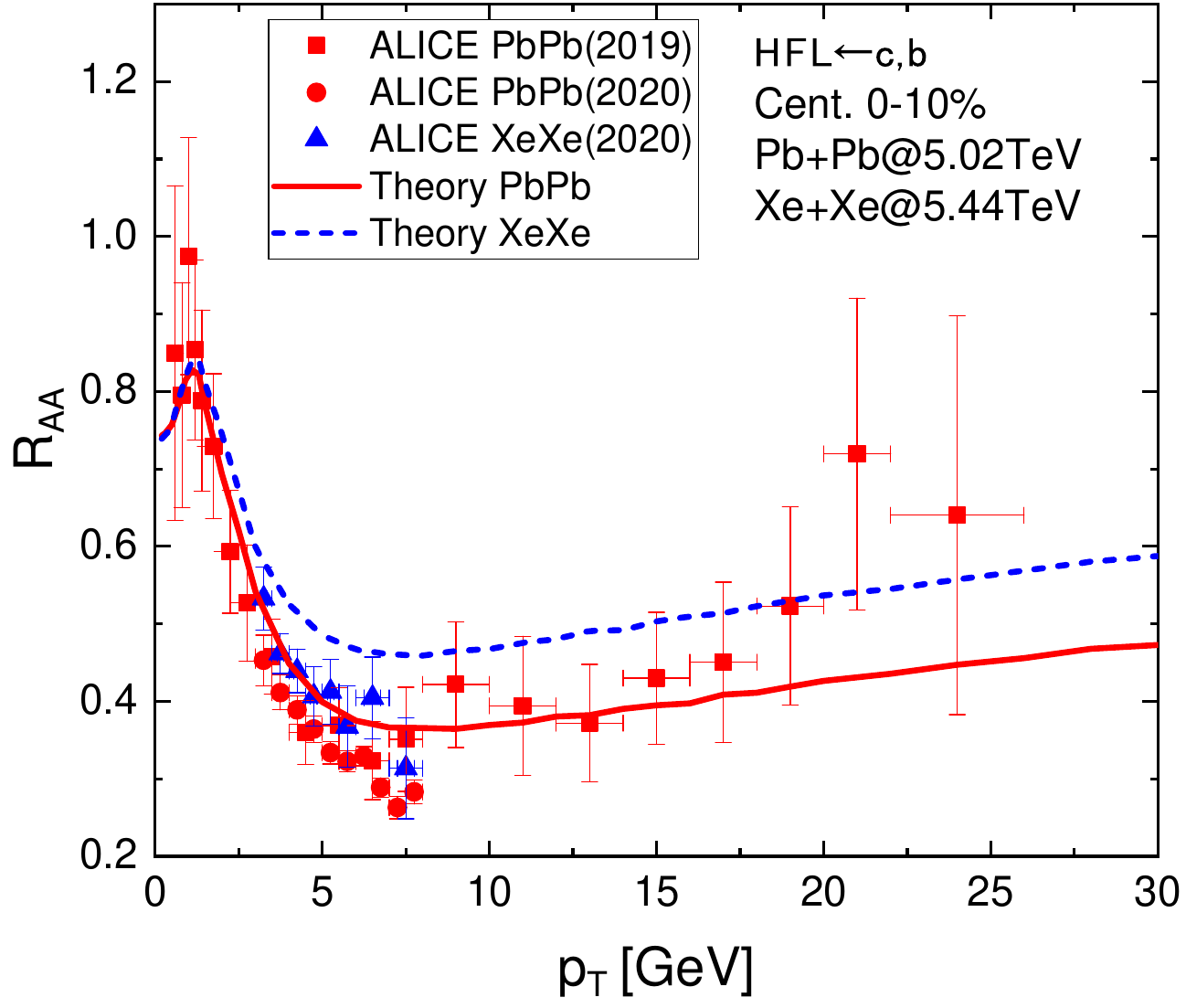}
\vspace*{0.1in}
\caption{The calculated $R_{AA}$ of HFL versus $p_T$ in central $0-10\%$ Xe+Xe collisions at $\sqrt{s_{NN}}=5.44$ TeV and Pb+Pb collisions at $\sqrt{s_{NN}}=5.02$ TeV compared to the ALICE data~\cite{ALICE:2019nuy,ALICE:2020try}.}
\label{fig:alixepb}
\end{center}
\end{figure}

\subsection{Path-length dependence of the HFL yield suppression in nucleus-nucleus collisions}
Besides exploring the mass effect of in-medium energy loss, the HFL production in high-energy nuclear collisions can also help probe the path-length dependence of the jet quenching effect in different collision systems \cite{Zigic:2018ovr, Shi:2018izg, Huss:2020dwe, Huss:2020whe, Zhang:2022fau, Li:2021xbd}. The colliding nucleus with a larger radius ($R_0$) is expected to form a medium with a larger size, in which the average path length of the jet propagation should be longer. As shown in Fig.~\ref{fig:alixepb}, we present the calculated $R_{AA}$ of HFL $\leftarrow c, b$ in central $0-10\%$ Xe+Xe collisions at $\sqrt{s_{NN}}=$ 5.44 TeV and Pb+Pb at $\sqrt{s_{NN}}=$ 5.02 TeV collisions compared to the ALICE data~\cite{ALICE:2019nuy,ALICE:2020try}. Within the same centrality bin, our theoretical results show stronger yield suppression of HFL $\leftarrow c, b$ in Pb+Pb collisions relative to that in Xe+Xe, which are generally consistent with the trend observed in the current ALICE measurements at $p_T<$ 8 GeV. Our calculations also predict that such distinction can be more significant at higher $p_T$, which will be interestingly tested with more precise measurements at the LHC.

\begin{figure}[!t]
\begin{center}
\vspace*{0.1in}
\includegraphics[width=3.3in,angle=0]{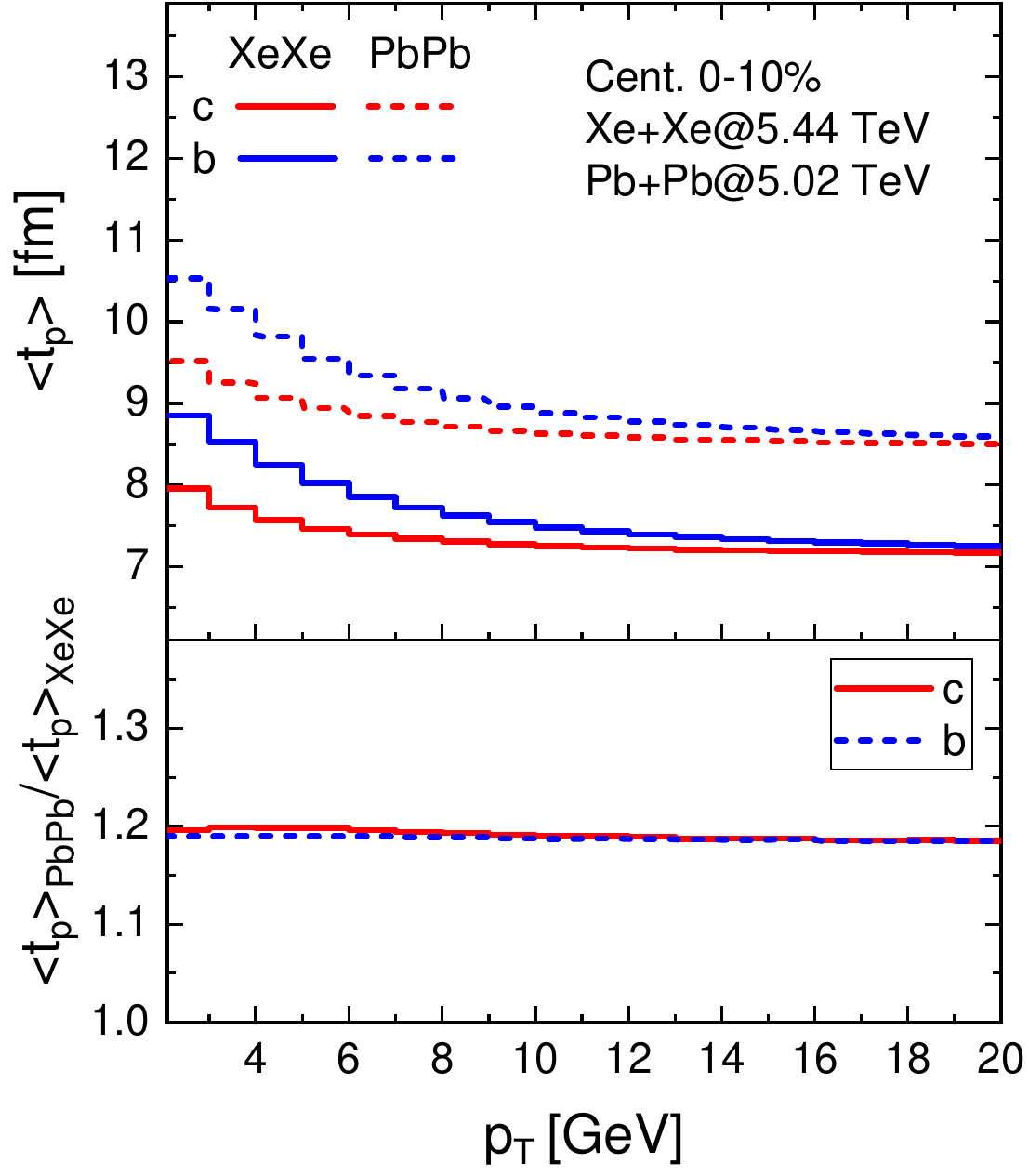}
\vspace*{-0.1in}
\caption{Mean propagation time $\left\langle t_p \right\rangle$ of heavy quarks (charm and bottom) as a function of their initial $p_T$ in central $0-10\%$ Xe+Xe collisions at $\sqrt{s_{NN}}=5.44$ TeV and Pb+Pb collisions at $\sqrt{s_{NN}}=5.02$ TeV, as well as the ratios of PbPb/XeXe in the lower panel.}
\label{fig:avetp}
\end{center}
\end{figure}

It is indisputable that the propagation time of heavy quarks in the QGP depends on the colliding nucleus's size and is closely related to their energy loss in A+A collisions. As shown in Fig.~\ref{fig:avetp}, we estimate the mean propagation time $\left\langle t_p \right\rangle$ of charm and bottom versus their initial $p_T$ in central $0-10\%$ Xe+Xe collisions at $\sqrt{s_{NN}}=5.44$ TeV and Pb+Pb collisions at $\sqrt{s_{NN}}=5.02$ TeV. Note that $t_p$ represents the propagation time of heavy quarks within the QGP phase ($T>T_c$) during the in-medium evolution in nucleus-nucleus collisions. It is found that, generally, heavy quarks with lower $p_T$ experience longer propagation time in the QGP due to their lower velocity. Similarly, for the same collision system, we also observe larger $\left\langle t_p \right\rangle$ of the bottom compared to charm due to the larger mass. From the comparison of two collision systems, the $\left\langle t_p \right\rangle$ in Pb+Pb collisions are larger than that in Xe+Xe collisions with the same centrality, consistent with our expectation that the Pb+Pb collisions may create a medium with larger size compared to Xe+Xe. The remarkable thing is that the ratio of $\left\langle t_p \right\rangle$ in Pb+Pb and Xe+Xe, about 1.2 as shown in the lower panel, is closely related to the radius ratio of the colliding nucleus ($R_0^{Pb}/R_0^{Xe}\sim$ 1.225) but independent of the flavour and initial $p_T$ of heavy quarks. The longer $\left\langle t_p \right\rangle$ of heavy quarks in Pb+Pb may play a key role in the stronger yield suppression of HFL $\leftarrow c, b$ compared to Xe+Xe.

\begin{figure}[!t]
\begin{center}
\vspace*{0.1in}
\includegraphics[width=3.3in,angle=0]{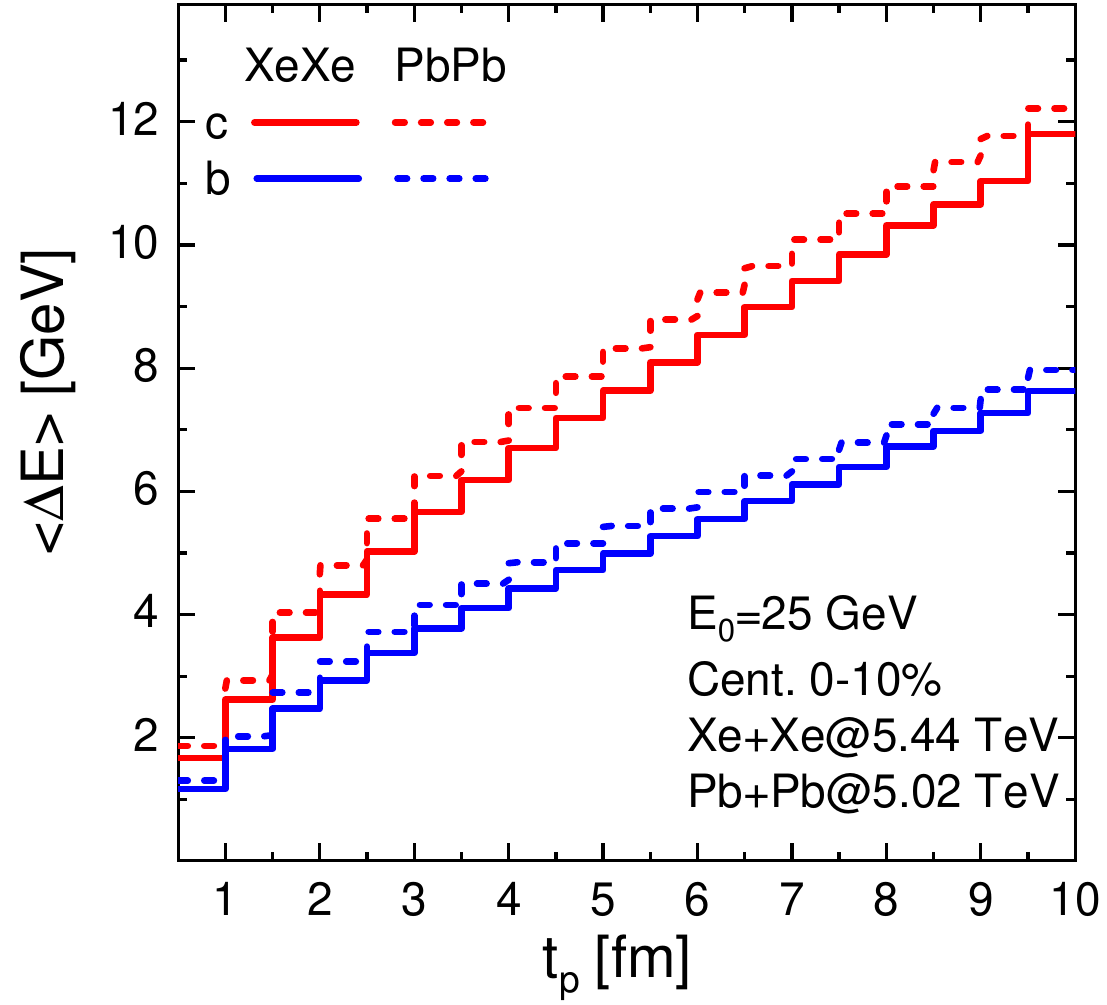}
\vspace*{-0.1in}
\caption{Mean energy loss $\left\langle \Delta E \right\rangle$ of heavy quarks versus propagation time $t_p$ in central $0-10\%$ Xe+Xe collisions at $\sqrt{s_{NN}}=5.44$ TeV and Pb+Pb collisions at $\sqrt{s_{NN}}=5.02$ TeV.}
\label{fig:avede}
\end{center}
\end{figure}

In Fig.~\ref{fig:avede}, we estimate the mean energy loss of heavy quarks (charm and bottom) with initial energy $E_0=25$ GeV as a function of propagation time $t_p$ in central $0-10\%$ Xe+Xe collisions at $\sqrt{s_{NN}}=5.44$ TeV and Pb+Pb collisions at $\sqrt{s_{NN}}=5.02$ TeV. It is clear to see that $\left\langle \Delta E \right\rangle$ increases with $t_p$, and at any $t_p$ charm quarks lose more energy than bottom quarks due to the ``dead-cone'' effect, no matter in Xe+Xe or Pb+Pb collisions. At the same time, we observe that $\left\langle \Delta E \right\rangle$ in Pb+Pb collisions is slightly larger than that in Xe+Xe with about $10\%$ at $t_p<4$ fm within the same centrality bin. It is because, within the same centrality bin, the QGP medium created in Pb+Pb collisions has a generally higher temperature than that in Xe+Xe. For instance, the temperature at the fireball's center is about 500 MeV in Pb+Pb while 480 MeV in Xe+Xe with $0-10\%$ centrality at $\tau_0=0.6$ fm \cite{Pang:2018zzo, Wu:2021fjf}. Since the strength of partonic interactions is closely related to the medium temperature ($\kappa\propto T^3$, $\hat{q}\propto T^3$) ~\cite{Kubo, Chen:2010te}, the higher temperature of the QGP medium formed in Pb+Pb leads to more effective energy loss of heavy quarks than that in Xe+Xe. From the above discussion, we conclude that the longer propagation time and more effective energy loss in Pb+Pb collisions play critical roles in the stronger yield suppression of the HFL compared to that in Xe+Xe.

\begin{figure}[!t]
\begin{center}
\vspace*{0.1in}
\includegraphics[width=3.2in,angle=0]{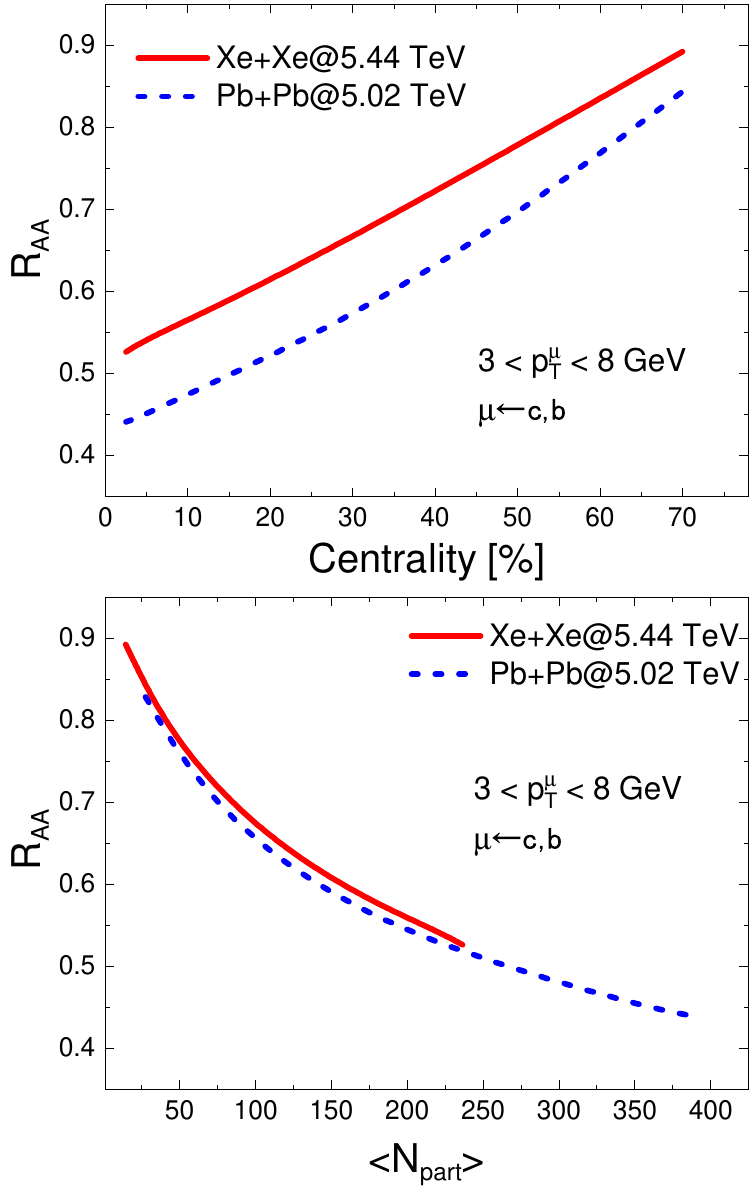}
\vspace*{-0.1in}
\caption{Integrated $R_{AA}$ of $\mu\leftarrow c, b$ from 3 to 8 GeV versus centrality (upper panel) and number of participants $\left \langle \rm N_{\rm part} \right \rangle$ (lower panel) both in Xe+Xe at $\sqrt{s_{NN}}=$ 5.44 TeV and Pb+Pb collisions at $\sqrt{s_{NN}}=$ 5.02 TeV.}
\label{fig:raacent}
\end{center}
\end{figure}

As shown in the upper panel of Fig.~\ref{fig:raacent}, we also calculate the integrated $R_{AA}$ of HFL $\leftarrow c, b$ within $3<p_T^{\mu}<8$ GeV as a function of centrality both in Xe+Xe and Pb+Pb collisions. As we can see, in the same centrality bin, the $R_{AA}$ of $\mu\leftarrow c,b$ in Pb+Pb collisions are lower than that in Xe+Xe, the same as the case in the most central collisions. However, as shown in the lower panel of Fig.~\ref{fig:raacent}, it is interesting to note that if one scales the $R_{AA}$ by the number of participants $\left \langle \rm N_{\rm part} \right \rangle$ instead of centrality, the differences of HFL $R_{AA}$ in Xe+Xe and Pb+Pb almost disappear at the overlap $\left \langle \rm N_{\rm part} \right \rangle$ region. Our findings are consistent with the previous study in Ref. \cite{Li:2021xbd}, in which the scaling behaviour of heavy-flavour hadron between different collision systems has been thoroughly discussed. The scaling behaviour of the HFL $R_{AA}$ indicates that the heavy quarks lose similar energy in the medium formed with the same $\left \langle \rm N_{\rm part} \right \rangle$ even for different collision systems.

\vspace{0.2in}
\section{Summary}
\label{sec:sum}
This paper presents a theoretical study of the lepton production from heavy flavour decay (HFL $\leftarrow c,b$) in high-energy nuclear collisions at the LHC. The initial $p_T$ spectra of heavy quarks are provided by the FONLL program, which matches the next-to-leading order hard QCD processes with the next-to-leading-log large-$p_T$ resummation. The in-medium propagation of heavy quarks is driven by the modified Langevin equations, which consider both the elastic and inelastic energy loss. The 2+1D CLVisc hydrodynamic model describes the space-time evolution of the fireball. Both the fragmentation and coalescence hadronization of heavy quarks are considered, and the semileptonic decay of heavy-flavour hadrons is employed by using PYTHIA8.

We present the calculated suppression factor $R_{AA}$ of HFL $\leftarrow c,b$ in Pb+Pb collisions at $\sqrt{s_{NN}}=5.02$ TeV compared to the ALICE and ATLAS data, which suggests lager $R_{AA}$ of HFL $\leftarrow b$ than that of HFL $\leftarrow c$. Furthermore, since the HFL $R_{AA}$ results in a complicated interplay of several factors, such as the pp-spectra, CNM effects, in-medium E-loss, fragmentation functions, coalescence hadronization, and decay channels, we propose a strategy to separate the individual influence of these factors to extract the net effect of mass-dependent E-loss. Based on quantitative analysis, we demonstrate that the coalescence hadronization, decay channels and the mass-dependent E-loss play an essential role at $p_T<5$ GeV, while the latter dominates the higher $p_T$ region. It is also found that the influences of CNM effects and FFs are insignificant, while different initial pp-spectra of charm and bottom quarks have a considerable impact at high $p_T$. In addition, we explore the path-length dependence of jet quenching by comparing the HFL $R_{AA}$ in two different collision systems: Xe+Xe at $\sqrt{s_{NN}}=5.44$ TeV and Pb+Pb at $\sqrt{s_{NN}}=5.02$ TeV. We obtain a smaller HFL $R_{AA}$ in Pb+Pb than that in Xe+Xe within the same centrality bin, consistent with the ALICE data. We find that the longer propagation time and more effective energy loss of heavy quarks in Pb+Pb collisions play critical roles in the stronger yield suppression of the HFL compared to that in Xe+Xe. At last, we observe a scaling behavior of the HFL $R_{AA}$ versus $\left \langle \rm N_{\rm part} \right \rangle$ in Pb+Pb and Xe+Xe collisions, which indicates that heavy quarks lose similar energy in the medium formed with the same $\left \langle \rm N_{\rm part} \right \rangle$ even for different collision systems.

However, we should be aware of the limitations of the phenomenological investigation presented in this work. Firstly, since the heavy quark energy loss at low $p_T$ region is inherently a non-perturbative effect, considering the diffusion coefficient obtained with the Lattice calculations should be helpful for understanding the low-$p_T$ heavy quark dynamics in heavy-ion collisions. In addition, note that in our current framework, the spatial diffusion coefficient $2\pi TD_s$ is assumed to be momentum-independent. In future studies, the temperature and momentum dependence of $D_s$ will be taken into account to consider the non-perturbative effects, which play an important role in the interaction between lower $p_T$ heavy quarks and the QGP medium.

{\it Acknowledgments:}
 This research is supported by the Guangdong Major Project of Basic and Applied Basic Research No. 2020B0301030008, and the National Natural Science Foundation of China with Project Nos.~11935007, 12035007 and 12247127. S. W. is supported by China Postdoctoral Science Foundation No. 2021M701279, the Talent Scientific Star-up Foundation (CTGU) No. 2024RCKJ013 and the Open Foundation of Key Laboratory of Quark and Lepton Physics (CCNU) No. QLPL2023P01.

\end{document}